\documentclass[conference]{misc/IEEEtran}
\newcommand{\added}[1]{{#1}}

\usepackage[hidelinks]{hyperref}
\usepackage{amsmath,amssymb,amsfonts}
\usepackage{algorithmic}
\usepackage{graphicx}
\usepackage{textcomp}
\usepackage[table]{xcolor}
\usepackage[utf8]{luainputenc}
\usepackage[british]{babel}
\usepackage[acronym, nomain]{glossaries}
\usepackage[capitalise]{cleveref}
\usepackage{marginnote}
\usepackage{tikz}
\usepackage{float}
\usepackage{tabularx}
\usepackage{rotating}

\usepackage{booktabs}

\def\BibTeX{{\rm B\kern-.05em{\sc i\kern-.025em b}\kern-.08em
T\kern-.1667em\lower.7ex\hbox{E}\kern-.125emX}}

\usetikzlibrary{patterns}
\usetikzlibrary{positioning}

\usepackage[backend=bibtex,style=ieee, mincitenames=1, maxcitenames=2, url=false, isbn=false]{biblatex}

\newacronym{ot}{OT}{Operational Technology}
\newacronym{it}{IT}{Information Technology}
\newacronym{cps}{CPS}{Cyber--Physical Systems}
\newacronym{stride}{STRIDE}{Spoofing, Tampering, Repudiation, Information
    Disclosure, Denial of Service, and Elevation of Privilege}
\newacronym{dfd}{DFD}{Data Flow Diagram}
\newacronym{iiot}{IIoT}{Industrial Internet of Things}
\newacronym{ics}{ICS}{Industrial Control System}
\newacronym{opcua}{OPC~UA}{Open Platform Communications Unified Automation}
\newacronym{osi}{OSI}{Open Systems Interconnection}
\newacronym{iso}{ISO}{International Organization for Standardization}
\newacronym{isa}{ISA}{International Society of Automation}
\newacronym{plc}{PLC}{Programmable Logic Controller}
\newacronym{tcp}{TCP}{Transmission Control Protocol}
\newacronym{udp}{UDP}{User Datagram Protocol}
\newacronym{ip}{IP}{Internet Protocol}
\newacronym{hmi}{HMI}{Human Machine Interface}
\newacronym{icmp}{ICMP}{Internet Control Message Protocol}
\newacronym{sctp}{SCTP}{Stream Control Transmission Protocol}
\newacronym{dhcp}{DHCP}{Dynamic Host Configuration Protocol}
\newacronym{tftp}{TFTP}{Trivial File Transfer Protocol}
\newacronym{rpc}{RPC}{Remote Procedure Call}
\newacronym{cvss}{CVSS}{Common Vulnerability Scoring System}
\newacronym{arp}{ARP}{Address Resolution Protocol}

\begin{document}

\title{
    AsIf: Asset Interface Analysis of Industrial Automation Devices
}

\author{\IEEEauthorblockN{Thomas Rosenstatter, Christian Schäfer, Olaf Saßnick,
    Stefan Huber}
  \IEEEauthorblockA{\textit{Josef Ressel Centre for Intelligent and Secure Industrial Automation} \\
    \textit{Salzburg University of Applied Sciences}\\
    Salzburg, Austria \\
    \{thomas.rosenstatter, christian.schaefer, olaf.sassnick, stefan.huber\}@ fh-salzburg.ac.at}
}

\maketitle

\begin{abstract}
As Industry 4.0 and the Industrial Internet of Things continue to advance, industrial control systems are increasingly adopting IT solutions, including communication standards and protocols. As these systems become more decentralized and interconnected, a critical need for enhanced security measures arises. Threat modeling is traditionally performed in structured brainstorming sessions involving domain and security experts. Such sessions, however, often fail to provide an exhaustive identification of assets and interfaces due to the lack of a systematic approach. This is a major issue, as it leads to poor threat modeling, resulting in insufficient mitigation strategies and, lastly, a flawed security architecture. 

We propose a method for the analysis of assets in industrial systems, with special focus on physical threats. Inspired by the ISO/OSI reference model, a systematic approach is introduced to help identify and classify asset interfaces. This results in an enriched system model of the asset, offering a comprehensive overview visually represented as an interface tree, thereby laying the foundation for subsequent threat modeling steps. To demonstrate the proposed method, the results of its application to a programmable logic controller (PLC) are presented. In support of this, a study involving a group of 12 security experts was conducted. Additionally, the study offers valuable insights into the experts' general perspectives and workflows on threat modeling.
\end{abstract}

\section{Introduction}
\label{sec:introduction}
\subsection{Motivation}
Due to the advancement of Industry 4.0, systems considered as \gls{ot}, such as
\glspl{ics} or \gls{iiot} are increasingly utilizing \gls{it} solutions. The
focus also shifts towards using open standards rather than proprietary
protocols. Prominent examples are communication protocols, such as IPv4/IPv6 or
\gls{opcua}. This leads to an increase of interconnectivity within shop level
machines and enterprise systems and also throughout the internet. With this
increasing connectivity also comes the need for advanced security. This is
especially relevant for \gls{cps}, as their physical interactions
can lead to severe consequences, such as damage to the environment, property or loss
of human lives~(safety)~\cite{zhu2011}.

However, the problem when applying traditional \gls{it} security solutions to
industrial systems is that they are not adapted to the unique requirements of
industrial systems. Stouffer et~al. list various points in~\cite{stouffer2022},
such as performance-, availability- or long spanning lifetimes criteria, which
are not always considered in such measures.

To address such concerns, international organizations have developed a variety
of standards and guidelines to enhance security in \gls{ot} environments. Two
of the most common ones are the IEC 62443 and NIST Special Publication 800--82.
However, these documents do not provide a step-by-step guide on how to
implement security, as each system is different and has its own, unique set of
requirements. Therefore, before designing a security architecture, security
experts must first understand the system, its assets, its functionality, and its
threats. For that, proper modeling of systems becomes paramount for all further
activities and can serve as a reference for security analysis later on.
However, the complexity of \gls{ot} systems makes modeling and understanding
them challenging, especially for non-security experts responsible for these
systems. Conversely, security experts may lack familiarity with \gls{ot}
systems and may not fully comprehend their specific requirements. 

\subsection{Problem Statement and Contribution}
Currently,
threat modeling processes for \gls{ot} systems often rely on structured brainstorming
sessions with experts from both domains. However, this approach may not be
sufficient, as important aspects might go unnoticed~\cite{messe2020}. This
non-exhaustive approach is not suitable and may lead to insufficient system
models. As a result, threat modeling efforts are often incomplete, lacking
in-depth analysis, and fail to capture all potential threat scenarios relevant
to the industrial systems.

These shortcomings further translate into deficient security architectures for
industrial systems~\cite{messe2020}. Inaccurate identification of risks and
vulnerabilities lowers the quality of security measures, leaving the systems
vulnerable to cyber-attacks. Without a solid security architecture, industrial
systems are susceptible to data breaches, operational disruptions, and
potential physical harm. Therefore, the foundation for a secure system is a
thorough understanding and modeling of the system and its assets.

In this paper, we aim to overcome the challenges of modeling \gls{ot} systems
and further conduct a study to gain insights into industry's current view on
threat modeling. More specifically, we contribute to the following research
questions: 

\begin{itemize}
    \item \textbf{RQ.1} How can we effectively address the challenges of
    exhaustively modeling \gls{ot} systems? 
    \item \textbf{RQ.2} What are the current industrial perspectives on threat
    modeling practices and their impact on security posture? 
\end{itemize}

We address \textit{RQ.1} by treating \gls{ot} systems foremostly as
distributed, cyber-physical systems. From this viewpoint, we propose
\textit{AsIf}, which is guided by the ISO/OSI model and helps with the
exhaustive identification and classification of the interfaces used by the
asset. \added{The proposed bottom-up approach, starting with the physical
interfaces allows for a thorough analysis in each layer enriching the
\textit{interface trees}. These trees visualize the system based on its
interfaces.}
For the evaluation,
we apply \textit{AsIf} to a real-world example in a industrial automation
testbed, namely a \gls{plc}.

Moreover, we conduct a study involving domain
experts to evaluate the methods utility \added{and superiority over their
currently used method.}
\added{In addition to the respondents evaluation of \textit{AsIf}, they} provide
valuable insights regarding their work in threat modeling (\textit{RQ.2}) and share our findings.

The remainder of the paper is organized as follows: First, a brief introduction
on threat modeling is given in \cref{sec:background-and-related-work}, followed
by the proposed method in \cref{sec:methodology}. The method is then applied to
a real-world example in \cref{sec:use-case}. An evaluation of the results and
discussion of the respondent's views on threat modeling is given in
\cref{sec:results-and-discussion}. Finally, the conclusion is presented in
\cref{sec:conclusion}.

\section{Related Work and Background}
\label{sec:background-and-related-work}
The concept of a \textit{dark factory} focuses on enabling fully autonomous production
operations through advanced automation and machine autonomy requiring four
design principles of Industry 4.0~\cite{hermann2016}: \textit{Interconnection},
\textit{Information Transparency}, \textit{Decentralized Decisions} and
\textit{Technical Assistance}. However, due to these principles and the change
of requirements, new cyber security challenges arise for industrial systems,
where traditional security measures are not sufficient enough~\cite{weiss2008}.
The term \textit{\gls{ot} security} is used to describe these differences and
challenges. As a crucial part in securing \gls{ot} system, system and threat
modeling processes must be adapted to the new requirements, too.

\subsection{Related Work}

Before threat modeling can be applied, the system must be analyzed,
broken down into its components and modeled in a way that enables the
identification of potential threats. This is the crucial first step, on which
all further processes are based and on which this paper puts the focus.

Shostack~\cite{shostack2014} recommends the use of \textit{software-centric}
approaches, as he includes the responsible software developers in the modeling
process. Using the documentation and the software code, a complete model of the
system is created. Hollerer et al.~\cite{hollerer2021} suggest using system
identification based on the ISA~95 network layout and using technical
documentation as reference material for modeling \gls{ot} systems. This
approach tackles the overall system with all its components and their
communication paths in the network. The MITRE ATT\&CK
framework 
supports the asset analysis
in the way that it provides a list of assets which is also mapped to threats,
albeit it does not provide means to systematically analyze the system in
question.

Khalil et al.~\cite{khalil2023} identified
assets by collaborating with system owners, developers and operators. They also
included various documentation and block diagrams describing the system. This
approach is similar to a brainstorming session, where all participants
contribute their knowledge and expertise.

However, the question is, whether these approaches are suitable for \gls{ot} systems.
They either consider the system as a complete white-box, which requires a
thorough understanding of the underlying software and needs the involvement of
the responsible developers. Or the assessment is done in a big-picture manner,
which potentially disregards important details of single components. Even
though the first approach is more thorough, it is not always possible to
involve the responsible persons or to get access to the source code. Especially
for legacy systems, this is often not possible. Therefore, a method is needed,
that combines the advantages of both approaches and substantiates its quality
with a comprehensive approach.

\subsection{Hybrid TCP/IP Model}

\begin{figure}[t]
    \centering
    \includegraphics[width=.8\linewidth]{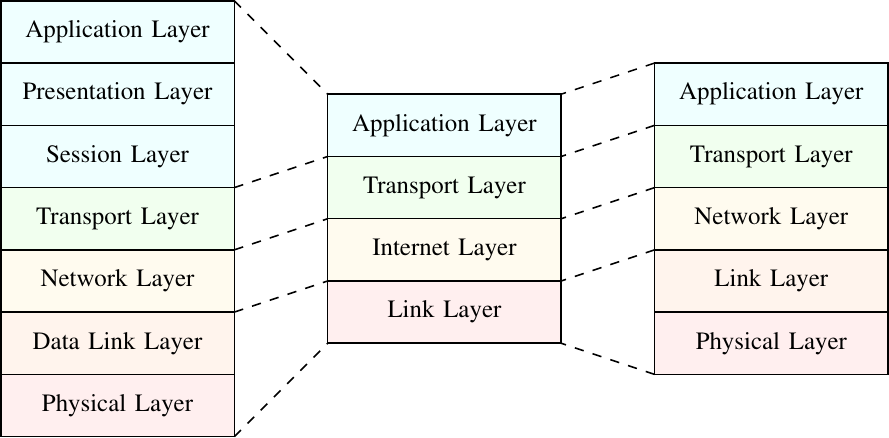}
    \caption{Comparison of the different network models: \gls{iso}/\gls{osi}
    model (left), the \gls{tcp}/\gls{ip} model (center) and Tanenbaum's hybrid
    TCP/IP model~\cite{tanenbaum2010} (right).}
    \label{fig:hybrid-tcp-ip-model}
\end{figure}

The %
TCP/IP model is a layered model, similar to the ISO/OSI model, used to
describe the communication between multiple systems. Compared to the 7-layer
ISO/OSI model, it comprises 4 layers: \textit{link}, \textit{internet},
\textit{transport} and \textit{application layer}. In contrast, the ISO/OSI
model contains the session and presentation layer, however, according to
Tanenbaum~\cite{tanenbaum2010}, these are of little use to most applications --
the functionalities are already implemented in the application layer. Regarding
the TCP/IP model, Tanenbaum criticizes the lack of a physical layer, which
describes the medium required for the transmission of bits, such as copper,
fiber or wireless. Although this may not be needed for the typical use of
TCP/IP, it is relevant in this work. Therefore, the 5-layered \textit{hybrid
TCP/IP} model by Tanenbaum~\cite{tanenbaum2010} is used in this work. It
separates the physical layer from the link layer, as shown in
Figure~\ref{fig:hybrid-tcp-ip-model}.

\subsection{Threat Modeling}

Once the system under consideration has been analyzed and a complete model can
be created, the next step is to identify potential threats. There are multiple
safety and security modeling methods available, each focusing on different
aspects: STPA-sec~\cite{young2014} focuses on system safety,
HAZOP~\cite{kletz1999} on hazards and system operability,
SAHARA~\cite{macher2015} on hazard, risk, and security,
PASTA~\cite{ucedavelez2015} on the process for attack simulation and
OCTAVE~\cite{alberts1999} on operationally critical threats and assets. 

In this work the \acrshort{stride} model is used in our demonstration, as it is lightweight
and provides a systematic way for modeling threats~\cite{khan2017}.
There is already research using \acrshort{stride} in industrial
domains~\cite{SRSH24}, indicating its applicability across various sectors. For
instance, domains such as smart-grid, Internet of Things, health-care and
automotive~\cite{khan2017,asif2021,omotosho2019,tuma2018b}.

Each letter in \acrshort{stride} represents a threat type, i.e., spoofing,
tampering, repudiation, information disclosure, denial of service and elevation
of privilege. \acrshort{stride} was developed by Microsoft and is a structured
approach to identifying potential threats to a system. The method requires an
already specified system architecture, its components, and their communication
paths. This modeling of the communication pathways can be achieved using
\glspl{dfd}. Once the system is fully modeled, STRIDE can be applied and each
component can be analyzed against each threat type.

Although our demonstration is based on \acrshort{stride}, we note that
\textit{AsIf} can be used in combination with any threat modeling technique
that follows some system model, such as a \gls{dfd}.

\section{Methodology}
\label{sec:methodology}

Besides the technical challenges of a \textit{dark factory} regarding automation
and autonomy aspects, security is a major concern in this context as well. 
To develop a proper security architecture, a thorough understanding of all systems 
and their potential interactions throughout their lifecycles is necessary.
Factories typically consist of systems manufactured by a
variety of manufacturers, who often provide sparse or no documentation for the entire
system internals. Adding to that, the documentation varies from manufacturer to manufacturer.
Therefore, a systematic and independent approach is required to assess the properties
of each system individually.

In our approach, the system is first broken down into its components and
analyzed individually. Once the individual requirements are defined, the overall
system can be discussed in a second step. This paper focuses on the first step,
the detailed analysis of the individual components.

The challenge we focus on is the security analysis of a single component. Our primary
concerns revolve around the interfaces of the device, whether they are physical
or logical, as these serve as potential entry points for attackers, in addition
to human errors and physical harm. There is a lack of guidance when modeling
the system's interfaces potentially leading to analyses based on an incomplete
model. Thus, a systematic approach to evaluate all interfaces is needed.

\begin{table}[t]
	\centering
	\caption{Methods for identifying interfaces and protocols.}
	\begin{tabularx}{\columnwidth}{cp{2.4cm}p{4.9cm}}
		\toprule \small
		\textbf{L.\#} & \textbf{Layer}             & \textbf{Exemplary Methods\textsuperscript{*}}                                 \\
		\midrule
		$5$ &Application Layer & Individually, based on the transport layer results from \texttt{Nmap} scan \added{and \texttt{tcpdump}}  for open TCP \& UDP ports                        \\
		$4$ &Transport Layer   & \texttt{Nmap} scan for supported IP protocols                                         \\
		$3$ &Network Layer     & \texttt{Wireshark} and \texttt{tcpdump} traffic analysis                  \\
		$2$ &Link Layer        & Derived from \added{Physical Layer}                        \\
		$1$ &Physical Layer    & Physical inspection                               \\
		\bottomrule
		\multicolumn{3}{l}{{\footnotesize{\added{\textsuperscript{*}Note, based on the environment, further methods may be useful.}}}}
	\end{tabularx}
	\label{tab:layer-analysis-suggestions}
\end{table}

\added{OT systems are distributed systems comprising several computing units
communicating with each other. These units each have their own communication
stack allowing the mapping to the ISO/OSI model. Therefore, we utilize the
hybrid TCP/IP model~\cite{tanenbaum2010} (see
Figure~\ref{fig:hybrid-tcp-ip-model}) as the starting point for the asset
interface analysis.} 
Compared to the classic TCP/IP model, we also need to model
physical interfaces, therefor we depend on the separation between physical and
network layer. In addition, modern protocols in OT are IP-based, hence the
ISO/OSI model would also be more general than needed. Layers 5-7 would also
require a deep understanding of the running applications, libraries, and parts
of the operating system. However, obtaining such exhaustive data on these
layers is impossible without access to the source code. Unlike other layers,
there is no definitive method to identify services or interfaces, often leading
to guesswork or reliance solely on documentation. The hybrid model consolidates
these software technologies into a unified application layer, simplifying the
focus on threat modeling for relevant applications. This approach is
advantageous for security experts, especially those unfamiliar with the target
device, as it streamlines the assessment process.

We propose to start the device analysis at the bottom and to then move upward
towards the application layer. For each layer, the interfaces and protocols used
are considered and the dependencies to the lower layers are noted. To achieve
this, we suggest individual identification methods for each layer, outlined in
Table~\ref{tab:layer-analysis-suggestions}. It should be noted that the
\texttt{Nmap} scans are technically performed one layer below than indicated,
however, the results of these scans are used for the layers as indicated in the
table. To apply the presented identification methods in a comprehensive manner, it may be necessary to consider the entire system lifecycle. 
For example, different services (i.e., interfaces) may be present in a maintenance or firmware update phase. 
Additionally, one needs to account for rare events, such as diagnostic data transmissions occurring only at fixed intervals.

This approach has the advantage of being systematic and comprehensive, as it
covers every network-related aspect of potential interfaces and data flows that
could be potentially exploited by malicious actors. The TCP/IP model describes the
internet architecture and as such does not provide options to model physical
interfaces, such as USB ports, DIP selectors and card slots for CF/SD
cards. To include use cases where operators interact with the hardware (by
pressing buttons, setting switches), we extend the model and add \textit{physical user
inputs and interactions} in the physical layer (L.1).

\begin{figure}[t]
    \centering
    \includegraphics[width=.6\linewidth]{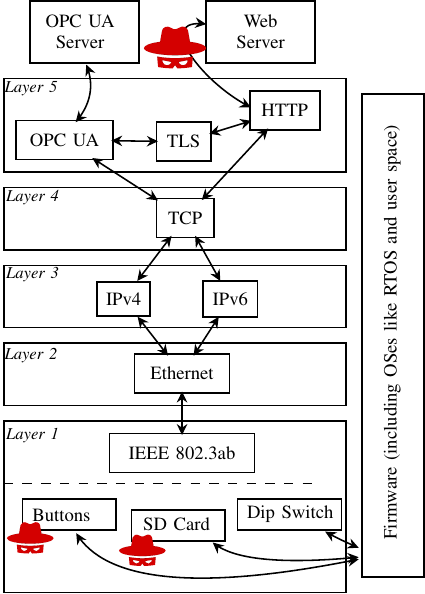}
    \caption{Extended TCP/IP model with example protocols. It shows
        the interfaces of an exemplary industrial device, which provides
        access via OPC UA and an HMI via a HTTP web server. As the
        firmware contains not only the applications, but also the software
        stacks and drivers for all other interfaces, it is modeled across all 
        layers. %
    }    \label{fig:interface-tree-example}
\end{figure}

By approaching the system in such a bottom-up manner, an \textit{interface tree} can
be created, providing information about all active interfaces and the
underlying services. This allows for a comprehensive overview of the system's
functionalities and enables an exhaustive modeling of the environment. Using
this model, \glspl{dfd} can be more easily derived for the threat modeling process. An
example of such an \textit{interface tree} is given in Figure~\ref{fig:interface-tree-example}. In this
figure, we also want to draw attention to the physical layer at the bottom,
which not only houses IEEE 802.3ab, but also other physical,
human-machine interfaces on the device. Other examples could be proprietary
network protocols or ports for serial communication. As these may not always
use upper layer services, we add a generic firmware application in parallel
to the model, to visualize the dependencies in a more understandable way.

\begin{figure*}[!htb]
    \centering
    \includegraphics[width=.8\linewidth]{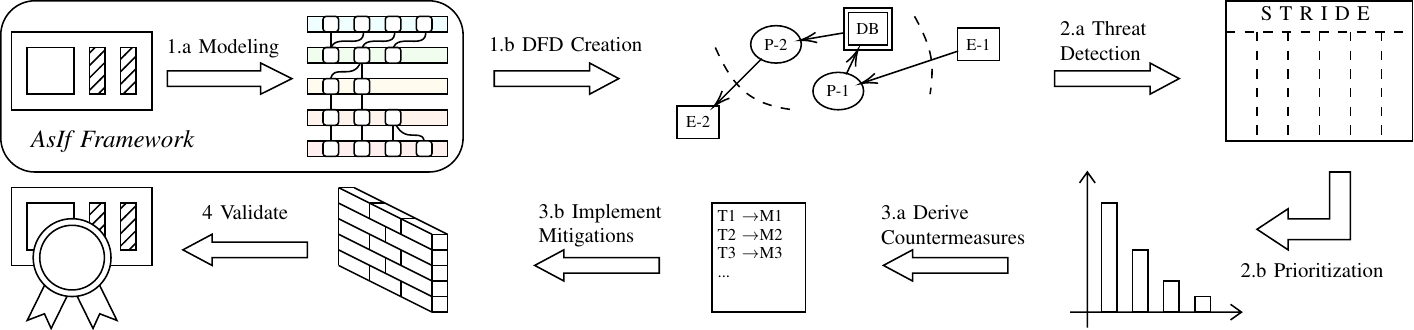}
    \caption{A threat modeling workflow, ranging from the system analysis and
        modeling (1.a) to the creation of a \gls{dfd} (1.b), the threat analysis
        (2.a) and prioritization (2.b), with \gls{stride} and \gls{cvss} exemplary,
        to the identification (3.a) and implementation (3.b) of countermeasures and
        lastly their validation (4). \added{The use of STRIDE is exemplary, any threat modeling method can be used based on \textit{AsIf}.}}
    \label{fig:threat-modeling-process}
\end{figure*}

\section{Use Case}
\label{sec:use-case}
To demonstrate the benefits of using \textit{AsIf}, we analyze
a \gls{plc}. As it is typically used in industrial automation and also provides
a wide range of physical, human-machine interfaces, it is a good example for
showcasing the use of our proposed approach.
Figure~\ref{fig:threat-modeling-process} illustrates in which step of the entire
threat modeling process \textit{AsIf} is applied. It also highlights the importance of a
thorough asset analysis during the modeling as all other steps rely on the
correctness and completeness of the provided information. Missing or incorrect
modeling could ultimately lead to an insecure system; for instance, interfaces
would not be considered.

We follow 
the steps outlined in Figure~\ref{fig:threat-modeling-process} until DFD creation, to highlight how this method can be applied
in practice. The focus lies \added{on the thorough identification and analysis of the interfaces following the proposed bottom-up approach} and the creation
of a \gls{dfd}. 

\subsection{Hardware}

The device under consideration features an ATOM 1.0 GHz processor, 256 MByte
DDR2 RAM, and a Compact Flash card slot for interchangeable program memory. The
device can be connected to a network via Ethernet or Powerlink, and provides a
RS232 interface for serial communication. Two modules for analog I/Os are
available, as well as a port for a proprietary fieldbus. There are multiple DIP
selectors, which can be used to set the network address and boot mode. Via a
button, a device reset can be initiated. To program and configure the \gls{plc}, a
proprietary software is used. This software and its documentation can be used
as sources of information for this analysis. The \gls{plc} was programmed with a
Windows~10 computer connected via Ethernet.

\subsection{Applying the Method}

For each layer of the model (see Figure~\ref{fig:hybrid-tcp-ip-model} and
Table~\ref{tab:layer-analysis-suggestions}), the interfaces are determined in a
manner, that is appropriate. The following describes the utilized methods and
lists the identified interfaces for each layer.

\textbf{Physical Layer.} Within this first, adapted, layer of the model, the
physical interfaces are described. Because of this adaption, not only network-
but also human-machine interfaces are encompassed. While these components are
not primarily used for communication, they present potential attack vectors
causing severe outcomes. As layer 1 deals with signal transmission technologies,
we argue that interfaces for human-machine interaction also fall in this layer,
as interacting with them involves electrical transmission of signals.
\cref{tab:layer-01-interfaces} shows all identified interfaces in the physical
layer of the \gls{plc}. The identification is done based on information from the
documentation and physical inspection of the device including the
inspection for possible interfaces behind the casing.

\begin{table}
    \renewcommand{\arraystretch}{1.2}
    \caption{Layer 1 Interfaces of the \gls{plc}. The top section shows
        classical network protocols. The lower section displays all other
        physical interfaces.}
    \footnotesize
    \begin{tabularx}{\columnwidth}{ll}
        \toprule
        \textbf{Interface}           & \textbf{Description}                                 \\
        \midrule
        IEEE 802.3ab        & 1000BASE-T Gbit/s Interface for Ethernet    \\
        IEEE 802.3u         & 100BASE-TX FE Interface for Powerlink      \\
        Fieldbus connectors& Interface for the proprietary used fieldbus \\
        RS232               & Serial connection with a programming device \\
        \midrule
        Analog I/O           & Read/write analog signals                          \\
        CF Card Slot        & Application storage                         \\
        USB Ports           & Can be used for USB peripherals             \\
        \added{RS232}               & \added{Serial interface}                            \\
        Button              & Resets the device                           \\
        DIP Selector        & Selects the boot mode                       \\
        2x DIP Selector     & Set the network address for programming     \\
        Extension Interface & Enables adding more extension modules       \\
        \bottomrule
    \end{tabularx}
    \label{tab:layer-01-interfaces}
\end{table}

\textbf{Link Layer.} The second layer contains the
communication protocols used on the physical layer. The \gls{plc}
supports Ethernet, Powerlink and a proprietary fieldbus.
Additionally, we argue, that USB and RS232 communication also fall in this
layer. The
information about the supported protocols is deduced from the previous layer
and the documentation.

\textbf{Network Layer.} The protocols for network-wide communication are
analyzed in this layer. Tools such as the network protocol analyzer
\texttt{Wireshark}\footnote{Wireshark: {https://www.wireshark.org/}} can be utilized for
sniffing the traffic at this layer and to derive the active protocols. The traffic
is captured during the boot process and normal operation to capture most active
communication processes. From this information, we were able to see that \gls{ip}v4, \gls{ip}v6 and ARP were used. Fieldbus and serial
communication protocols do not have routing capabilities, so they do not exist
in this layer and beyond.

\textbf{Transport Layer.} At this layer, the documentation did not provide any
more information about which protocols besides \gls{tcp} and the UDP are
supported or active. The network mapper \texttt{Nmap}\footnote{Nmap:
\url{https://nmap.org/}} is chosen as it allows to run an \gls{ip} scan, which
exhaustively iterates through all possible \gls{ip} protocol numbers. Depending on the
responses \texttt{Nmap} receives, it interprets the protocols that are supported
by the host\footnote{\added{command: \texttt{nmap -p0- -sO -T4 [ip address]}}}. In this use case, it was revealed that the \gls{plc} not only
supports \gls{tcp}, \gls{udp} and ICMP, but also SCTP. 

\begin{figure*}[ht!]
    \centering
    \includegraphics[width=\textwidth]{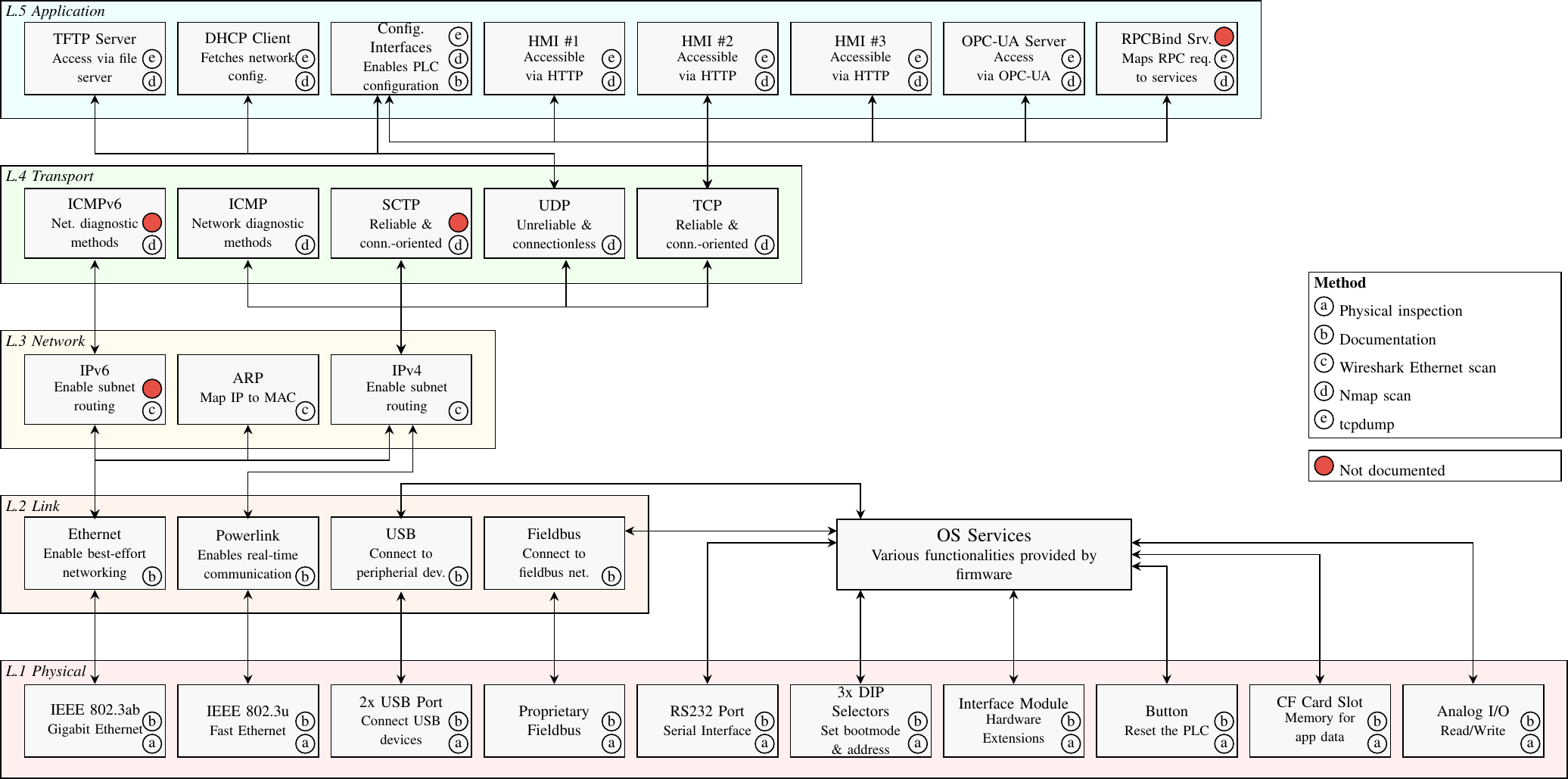}
    \caption{Resulting interface tree after applying the \textit{AsIf} method on the \gls{plc}.}
    \label{fig:x20-interface-tree-graph}

\end{figure*}

\textbf{Application Layer.} Networked applications typically operate utilizing
ports for communication, thus the \gls{plc} is scanned for open ports using
\texttt{Nmap}. This scan provides insights about active services running on
the system. The scans\footnote{\added{command: \texttt{nmap -p0- -s[S|U] -T4 [ip address]}}} are performed on all 65535 \gls{tcp} and \gls{udp} ports.
The internal port-service mapping of \texttt{Nmap} may not be correct for all
services as some manufacturers may use proprietary protocols. Hence, we
recommend a manual verification.

The port scan in combination with the programming software and
its documentation may help to further identify the running services on the
\gls{plc}. In this use case, multiple web servers acting as \glspl{hmi}, an
\gls{opcua} server, a TFTP server, a DHCP client, a RPCbind
server and a remote configuration service are identified.

Once all the interfaces on the target system are known, a model can be created to
visualize the dependencies across all layers. The model, the \textit{interface tree} follows the architecture of the hybrid
\gls{tcp}/\gls{ip} model and can help security experts to get a better
understanding of the systems internals. Figure~\ref{fig:x20-interface-tree-graph}
shows the \textit{interface tree} for the investigated \gls{plc}. The Figure also highlights the protocols that were found during the assessment, but which were not documented.

\subsection{Generating a Data Flow Diagram}

When constructing \gls{dfd}s %
it is essential
to focus on two layers: the application layer and the physical layer.
Firstly, by analyzing running applications within the system, associated
processes and data flows can be derived. These are ultimately responsible for
communication both within and beyond the system, making them primary targets
for potential security threats. Secondly, the physical interfaces require
special attention, as they enable human-machine interactions independent of the
network and thus allow cybersecurity measures to be bypassed. For instance, an
adversary could disguise as a service technician and potentially be able to substitute the CF card containing malicious firmware or exploit
the serial interface to reprogram the device.

An important aspect of industrial devices are the different phases of their
lifecycle, as they may have different requirements for security. For instance,
during the installation and maintenance phases, access must be allowed for
configuration tools to obtain the device. Security measures must also consider these situations. On the other hand, during operation, these interfaces
should be disabled. Therefore, we chose to draw multiple diagrams, each
depicting a different phase of the lifecycle. This way, the diagrams are not
too complex and additional awareness for the different phases is raised. 

\begin{figure*}[t]
    \centering
        \includegraphics*[width=.9\linewidth]{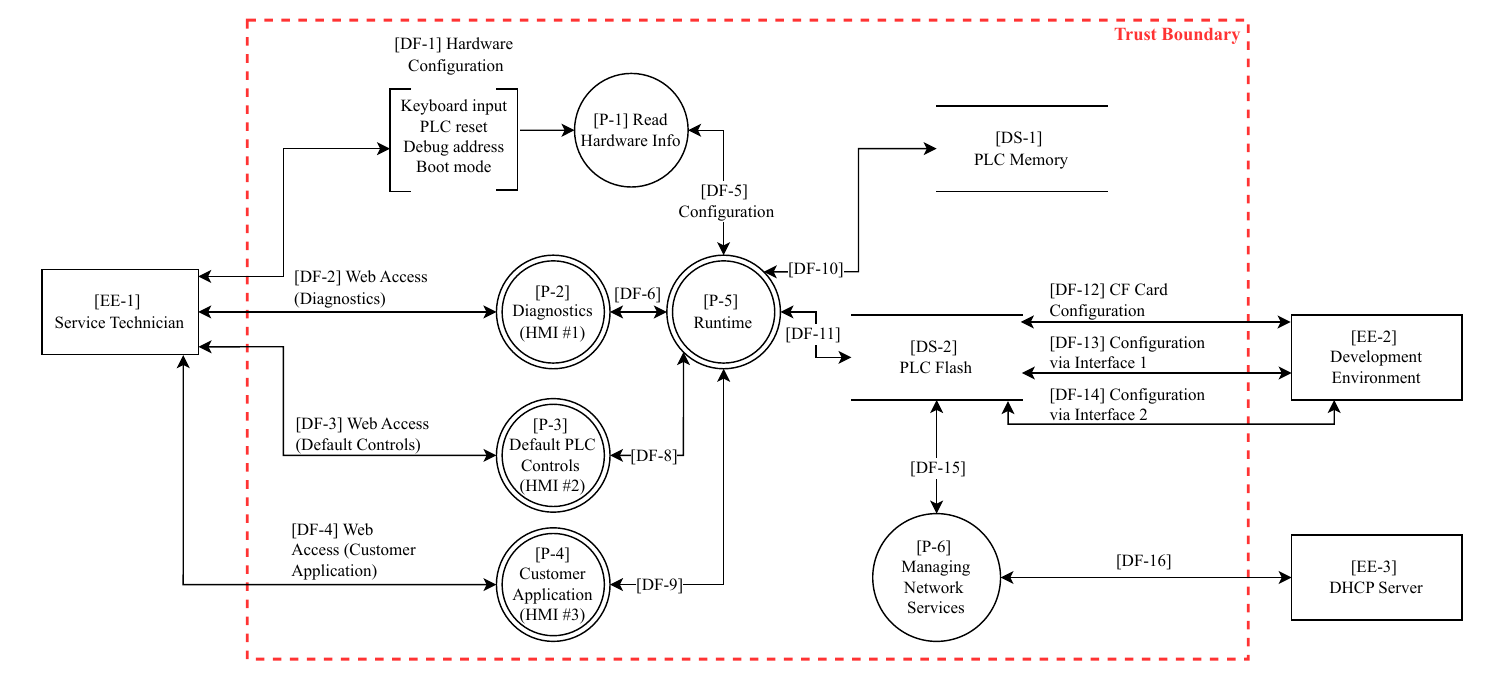}
    \caption{Data flow diagram of the \gls{plc} during the service phase.}
    \label{fig:x20-dfd}
\end{figure*}

For
the described use case, a \gls{dfd} of the \gls{plc} during a generic
`service' phase is depicted in Figure~\ref{fig:x20-dfd}. The service phase includes configuration,
maintenance and update activities. %
The \gls{dfd} illustrates three external entities (EE-1 to EE-3) interacting
with the \gls{plc}. The hardware configuration (DF-1) uses input through
physical interfaces (see Figure~\ref{fig:x20-interface-tree-graph}) such as the
USB-connected keyboard, DIP switches and a button. The access to the web
services is provided through the three identified HMIs, which in turn interact
with the runtime. On the right side, the DHCP server is shown, as it is used for
receiving a valid IP address and further network information (DF-16). Last, the
development environment may interact via two configuration interfaces (L.5)
provided by either TCP or UDP (L.4). In addition, the service technician may
also manipulate or replace the CF card (L.1), thus also altering the system. 

Figure~\ref{fig:x20-dfd} highlights %
the importance of a systematic analysis as we propose, as otherwise relevant data 
flows simply could have been forgotten which ultimately might result in a 
security incident halting the entire production.

\section{Evaluation}
\label{sec:results-and-discussion}

The practicality and benefits of using the \textit{AsIf} framework for
identifying assets through interface analysis in a structured manner was
evaluated by domain experts. The details of the study and its results are
analyzed and presented below. The framework was primarily perceived as useful
and considered to improve the existing processes the experts are currently
using. 7 out of 11 respondents consider to implement the proposed framework. In
the following sections the design of the study through a questionnaire is
discussed, and the experts' general view on the need for threat modeling and its
challenges (\textit{RQ.2}) are presented. Thereafter the feedback to the
\textit{AsIf} framework is discussed in more detail. 

\subsection{Design of the Study}\label{subsec:design:questionnaire} The
evaluation is designed to address security experts working in automation or
other companies within this industry (e.g., security consultant and
researchers). It is explicitly targeted at employees from R\&D
departments or similar. The aim is to get direct feedback from experts in closer
proximity, thus the evaluation material is provided in German. 
Given that all participants in the study are working in the Central European region,
the results might be biased toward the prevailing regional working culture.
The material consists of an introductory video\footnote{Video: 
\url{https://www.youtube.com/watch?v=2GpFI3XDmgA}
} and a follow up
questionnaire\footnote{Questionnaire and presentation:
\url{https://doi.org/10.5281/zenodo.11201810}
}.
First, questions to the background of the respondent are asked, e.g., domain,
and experience in safety and security. These questions are followed by
investigating the respondent's perception on the need for threat modeling and
the challenges they face. These insights are used to contribute to \textit{RQ.2}.
Last, questions about our proposed framework are asked. 

We received 12 responses to the questionnaire with the majority (75\% resp.~9)
working in organizations with more than 250 employees. Furthermore, the majority
of the respondents~(7) have worked in cybersecurity for five or more years.
Their safety background is more limited. Seven of the respondents have
experience in safety for less than 2 years. Despite having received responses
from only 12 participants, this initial feedback is highly valuable for
evaluating the framework. Particularly, because the respondents are experts in
the targeted domain. Their specialized insights ensure that even this small set
of people provides a robust foundation for assessing the utility of the
framework and its further development. \subsection{Industry's Perception on
Threat Modeling}\label{subsec:industry:perception}

\begin{figure}[t]
    \centering
    \includegraphics[width=\linewidth]{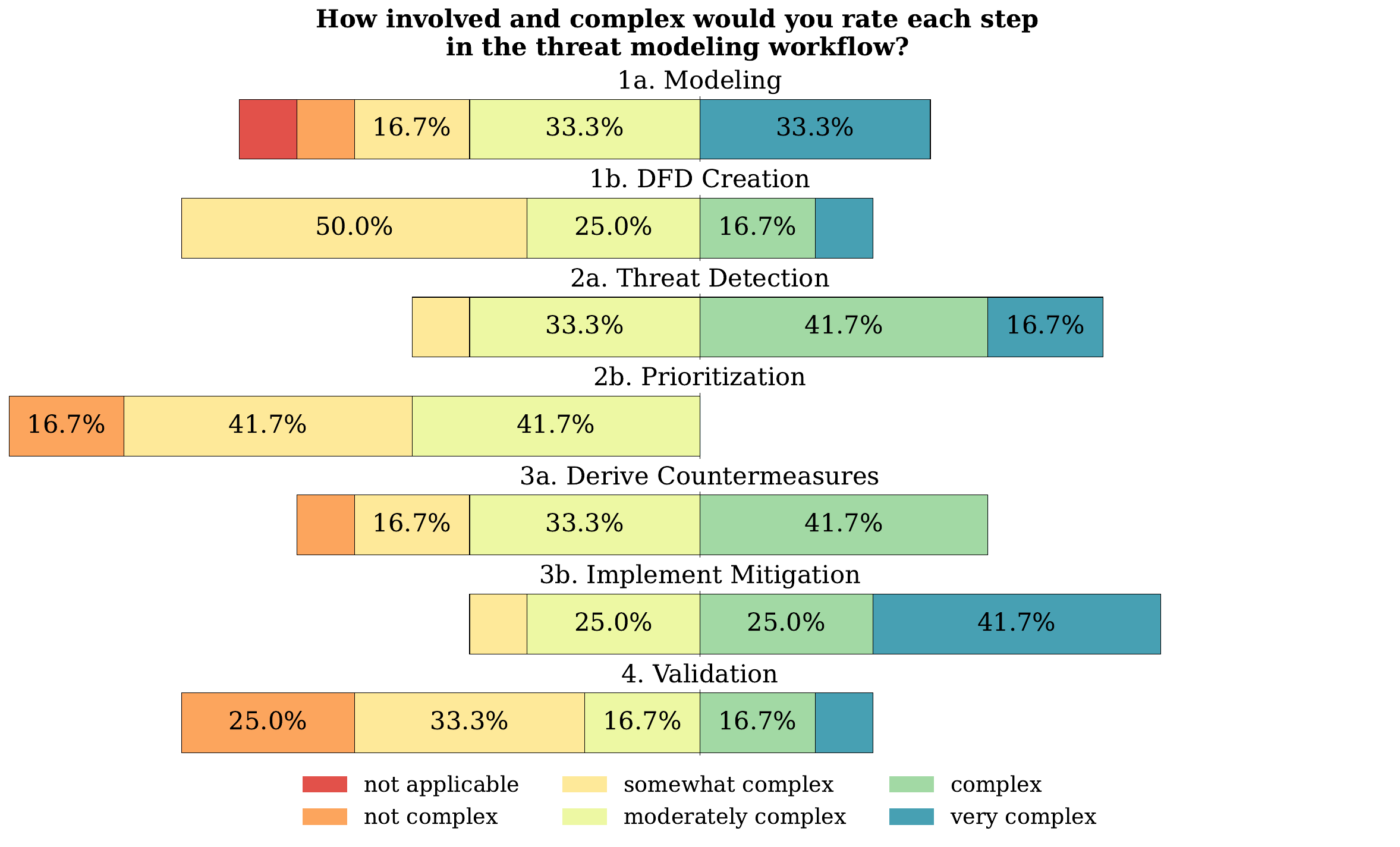}
    \caption{Response to the question on how involved and complex the experts
    perceive each step in the presented threat modeling workflow
    (n=12).}\label{fig:questionnaire:complexity}
\end{figure}

Following the steps of threat modeling as illustrated in
Figure~\ref{fig:threat-modeling-process}, the participants were asked about the
perceived complexity and involvement required for each step. As shown in 
Figure~\ref{fig:questionnaire:complexity}, the steps modeling, threat detection, and
implementing mitigation techniques were generally regarded as involved and
complex. Prioritization and validation on the other hand, were considered
to be the least involved and complex. 

When asked about the most challenging topics, nine
participants stated that the identification of attack vectors is challenging or
very challenging. Building knowledge about current security topics was also
considered by more than half as challenging or very challenging. This is
probably caused by the fast pace at which cybersecurity evolves. Moreover, having a
systematic approach and comprehensible documentation was considered challenging
by a third of the respondents. 

The challenges to have a systematic approach for threat modeling and the
involvement modeling requires is specifically what the \textit{AsIf} framework
aims to provide support. However, the questionnaire also shows some other
aspects in which experts would benefit from support. For instance, one
respondent argued that more practical approaches are needed. Several respondents
see traceable documentation and automation of the threat analysis as the biggest
areas for improvement. 
While the proposed \textit{AsIf} framework does not directly address automation, 
thereby improving the scalability, some of the individual identification methods 
in Table~\ref{tab:layer-analysis-suggestions} can be carried out in an automated manner.

\subsection{Framework Evaluation}\label{subsec:framework:eval} The last part of
the evaluation consisted of six questions about the framework. Five questions
could each be answered on a scale ranging from $1$ corresponding to \textit{no}
to $5$ corresponding to \textit{yes}. A free text question was added last to get
further remarks on the framework. 

\begin{figure}[t]
    \centering
    \includegraphics[width=.9\linewidth]{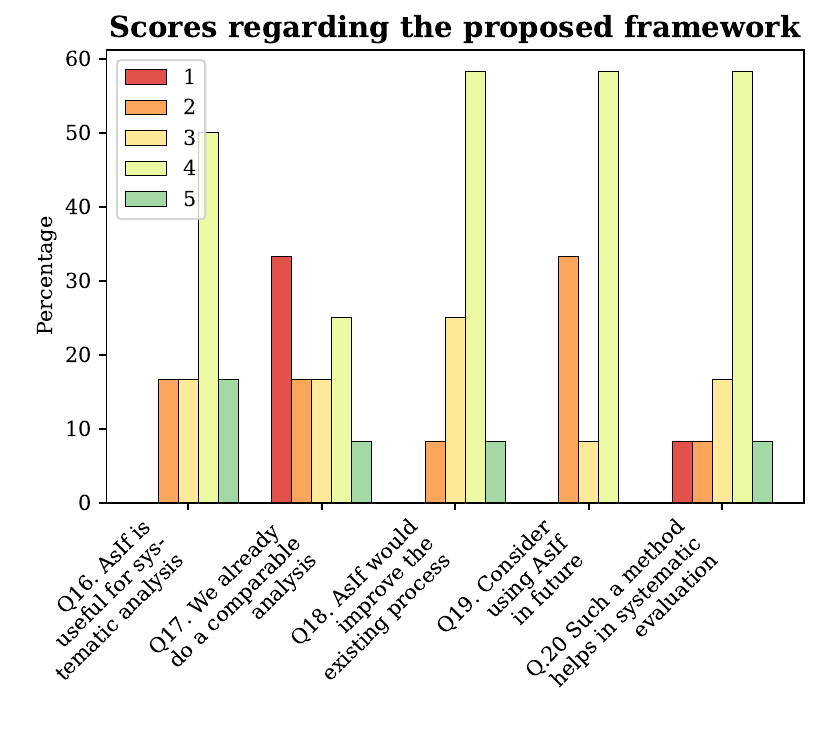}
    \caption{Responses to the \textit{AsIf} framework. Each question (Q\#) could be
    answered on a scale ranging from $1$/\textit{no} to
    $5$/\textit{yes}.}\label{fig:feedback:asif}
\end{figure}

Figure~\ref{fig:feedback:asif} summarizes the response of the first five
questions\footnote{Note that each question was originally formulated as
question. They are only shortened for illustrational purpose. For the precise
formulation see the additional material 
\url{https://doi.org/10.5281/zenodo.11201810}
.}. The overall response to the
framework is very positive. \textit{AsIf} is considered as a useful tool for
systematic analysis (Q16) and would improve the existing processes for asset
identification (Q18). Furthermore, more than 50\% (who rated $4$) would consider
using this methodology (Q19) underlining the practicality of the framework. The
majority also believes that \textit{AsIf} or a comparable approach would help
them in the systematic evaluation of their system (Q20). Interestingly, the
answers to the question (Q17) whether the respondents already use a comparable
analysis was answered quite mixed, showing no specific trend. Four participants
even answered with a clear \textit{no}~(rated $1$).

We only received two full text comments. The first is addressing the lack of
automation in \textit{AsIf}. We agree that the method, as it is proposed, is
difficult to automate, since it requires the manual investigation of the system
under consideration throughout all layers. The second comment is acknowledging
that it is especially useful when analyzing existing systems, however, the respondent also highlights that
\textit{AsIf} would not add additional support when designing the system, 
as the architecture, interfaces and functions are defined during the design
phase. The same respondent also added that the proposed illustration in form of
an \textit{interface trees} is a great supporting tool to graphically represent the
connections and discuss them in the team. 

\section{Conclusion}
\label{sec:conclusion}
A comprehensive asset analysis is essential to model the system in sufficient
detail, as this represents the initial step in threat modeling. Given that
operational technology systems are distributed cyber-physical systems, the
proposed \textit{AsIf} framework is inspired by the TCP/IP model allowing for a
systematic analysis of the interfaces across all communication layers. \added{By
introducing} \textit{interface trees} for modeling and visualization,
\textit{AsIf} enhances traceability of interfaces throughout all layers.
\added{This approach also reinforces the consideration of lifecycles, as, for
instance, some physical interfaces are only required in specific operation modes.}

The evaluation of the framework is twofold: initially, a use case demonstrated
the application of \textit{AsIf}, followed by a study to gather feedback from
security experts in the industry. The positive response not only underscores the
effectiveness of \textit{AsIf} in enhancing evaluation processes but also
indicates a strong inclination to adopt similar methodologies in the future.

In addition to the study, we engaged with industry experts to collect their
insights on threat modeling, aiming to identify critical areas for %
further research.

 \section*{Acknowledgment} 

 The financial support by the Christian Doppler Research Association, the
 Austrian Federal Ministry for Digital and Economic Affairs and the Federal State
 of Salzburg is gratefully acknowledged.

\medskip
\renewcommand*{\bibfont}{\footnotesize}
\printbibliography

\end{document}